# Two characteristic volumes in thermal nuclear multifragmentation

V. A. Karnaukhov (1)[*], H. Oeschler (2), S. P. Avdeyev (1), V. K. Rodionov (1), V. V. Kirakosyan (1), A. V. Simonenko (1), P. A. Rukoyatkin (1), A. Budzanowski (3), W. Karcz (3), I. Skwirczyńska (3), E. A. Kuzmin (4), L. V. Chulkov (4), E. Norbeck (5), A. S. Botvina (6)

[1] *Joint Institute for Nuclear Research, Dubna, Russia*

[2] *Institut für Kernphysik, Darmstadt University of Technology, Darmstadt, Germany*

[3] *H.Niewodniczanski Institute of Nuclear Physics, Cracow, Poland*

[4] *Kurchatov Institute, Moscow, Russia*

[5] *University of Iowa, Iowa City, USA*

[6] *Institute for Nuclear Research, Moscow, Russia*

The paper is devoted to the experimental determination of the space-time characteristics for the target multifragmentation in $p(8.1\text{GeV}) + \text{Au}$ collisions. The experimental data on the fragment charge distribution and kinetic energy spectra are analyzed within the framework of the statistical multifragmentation model. It is found that the partition of hot nuclei is specified after expansion of the target spectator to a volume equal to $V_t = (2.9 \pm 0.2)\, V_o$, with $V_o$ as the volume at normal density. However, the freeze-out volume is found to be $V_f = (11 \pm 3)\, V_o$. At freeze-out, all the fragments are well separated and only the Coulomb force should be taken into account. The results are in accordance with a scenario of spinodal disintegration of hot nuclei.

## 1. THERMAL MULTIFRAGMENTATION AND NUCLEAR FOG

The study of the decay of very excited nuclei is one of the most challenging topics of nuclear physics giving access to the nuclear equation of state for the temperatures below $T_c$ – the critical temperature for the liquid-gas phase transition. The main decay mode of very excited nuclei is a copious emission of intermediate mass fragments (IMF), which are heavier than $\alpha$-particles but lighter than fission fragments. The great activity in this field has been stimulated by the expectation that this process is related to a phase transition in nuclear media.

---

[*]Email address: karna@nusun.jinr.ru

An effective way to produce hot nuclei is reactions induced by heavy ions with energies up to hundreds of MeV per nucleon. Around a dozen sophisticated experimental devices were created to study nuclear multifragmentation with heavy ion beams. But in this case the heating of the nuclei is accompanied by compression, strong rotation, and shape distortion, which may essentially influence the decay properties of hot nuclei. Investigation of dynamic effects caused by excitation of collective degrees of freedom is interesting in itself, but there is still the problem of disentangling all these effects to extract information on the thermodynamic properties of the hot nuclear system.

One gains simplicity, and the picture becomes clearer, when light relativistic projectiles (first of all protons, antiprotons, pions) are used. In this case, in contrast to heavy ion collisions, fragments are emitted by only one source – the slowly moving target spectator. Furthermore, its excitation energy is almost entirely thermal. Light relativistic projectiles provide therefore a unique possibility for investigating *thermal multifragmentation*, which was realized in the projects ISiS (Bloomington, Indiana), MULTI (Kyoto, Tokyo, Tsukuba) and FASA (Dubna). The decay properties of hot nuclei are well described by statistical models of multifragmentation [1,2] and this can be considered as an indication that the system is thermally equilibrated or, at least, close to that. For the case of peripheral heavy ion collisions the partition of the excited system is also governed by heating.

In several papers, multifragmentation of hot nuclei is considered as spinodal decomposition. The appearance of the unstable spinodal region in the phase diagram of nucleonic system, as like as for the classical one, is a consequence of similarity between nucleon-nucleon and Van der Waals interactions [3-5]. The equations of state are very similar for these systems, which are very different in respect to the size and energy scales.

One can imagine that a hot nucleus (at $T$ = 5-7 MeV) expands due to thermal pressure and enters the unstable region. Due to density fluctuations, a homogeneous system is converted into a mixed phase consisting of droplets (IMF) and nuclear gas (nucleons and light clusters with $Z \leq 2$) interspersed between the fragments. Thus the final state of this transition is a *nuclear fog* [3], which explodes due to Coulomb repulsion and is detected as multifragmentation. Therefore, it is more appropriate to associate the spinodal decomposition with the *liquid-fog* phase transition in a nuclear system rather than with the *liquid-gas* transition, as stated in several papers (see for example [6-8]).

The phase transition scenario is evidenced by number of observations, some of them are the following:

   a) density of the system at the break-up is much lower compared to the normal one;

   b) mean life-time of the fragmenting system is very small ($\approx$ 50 *fm/c*), which is in the order of the time scale of density fluctuation (see for example [9]);

   c) break-up temperature is lower than $T_c$, the critical temperature for the *liquid-gas* phase transition [10].

The first point from this list requires a more detailed experimental study. There are a number of papers with estimations of the characteristic volume (or mean density) by analysis of different observables in multifragmentation. The values obtained deviate significantly. For example, the mean freeze-out density of about $\rho_0/7$ was found in Ref. [11] from the average relative velocity of IMFs at large correlation angles for $^4$He + Au collisions at 1.0 and 3.6 GeV/nucleon using the statistical model MMMC [2]. In paper [12] the nuclear caloric curves was considered within an expanding Fermi gas model to extract average nuclear densities for different fragmenting systems. It was found to be ~0.4 $\rho_0$ for medium and heavy masses. In Ref. [13] the mean kinetic energies of the detected fragments are analysed by applying energy balance, calorimetric measurements and calculations of many-body Coulomb trajectories. The freeze-out volume is found to be ~ $3V_0$ for the projectile fragmentation in Au+Au collisions at 35 MeV per nucleon ($V_0$ is the source volume at normal density). The statistical model of multifragmentation (SMM) [1] has been used in this analysis. In a recent publication [7] the average source density for the fragmentation in the 8.0 GeV/*c* $\pi^-$ + Au interaction is estimated to be ~ $(0.25-0.30)\rho_0$ at $E^*/A \sim 5$ MeV from the moving-source-fit Coulomb parameters.

In this paper we extracted the characteristic volumes of the fragmenting system produced by *p*(8.1GeV) + Au collisions by two methods: i) by analyzing the IMF charge distribution; ii) from the shape of the kinetic energy spectra of carbon. It is done within SMM. Very different values of volumes are obtained by these two methods. The meaning of this observation is discussed. Experimental data have been obtained using the FASA device installed at the external beam of the Synchrotron-Nuclotron accelerator complex (Dubna).

## 2. VOLUME FROM IMF CHARGE DISTRIBUTIONS

The shape of the IMF charge distribution depends on the size of the system at the moment of partition, as demonstrated in our paper [10] for fragmentation in $p(8.1\text{GeV})+\text{Au}$ collisions. The reaction mechanism for light relativistic projectiles is usually divided into two stages. The first one is a fast energy-depositing stage during which very energetic light particles are emitted and the target spectator is excited. We use the intranuclear cascade model (INC) from Ref. [14] for describing the first stage. The second stage is considered within the framework of the statistical model of multifragmentation (SMM), which describes multibody decay (volume emission) of a hot and expanded nucleus. But such a two-stage approximation fails to predict the measured fragment multiplicity. To overcome this difficulty, an expansion stage is inserted (in spirit of EES model [15]). The residual (after INC) masses and their excitation energies are tuned (on event-by-event basis) to obtain agreement with the measured mean IMF multiplicity (see [16] for details). We call this combined model the INC+ Exp.+SMM approach.

The break-up (or partition) volume is parameterized in the SMM as $V = (1+k) V_o$. It is assumed in the model that the freeze-out volume, defining the total Coulomb energy of the final channel, coincides in size with the system volume at the moment of partition. Thus, $k$ is one of the key parameters of SMM, which also defines (in the first approximation) the free volume ($\approx kV_o$) and contribution of the translation motion of fragments to the entropy of the final state. Within this model the probabilities of different decay channels are proportional to their statistical weights (exponentials of entropy). The entropy is calculated using the liquid-drop model for hot fragments. The statistical model considers the secondary disintegration of the excited fragments to get the final charge distribution of cold IMFs. The importance of the secondary decay stage was analysed in Ref. [9].

Figure 1 shows the IMF charge distribution for $p$ (8.1GeV) + Au collisions measured at $\theta = 87°$. Error bars do not exceed the symbol size. Fragment yields are corrected for the detection efficiency, which is slightly charge dependent. The lines are obtained by calculations using the INC+Exp.+SMM prescription under three assumptions about the fragmenting system volume $V$: it is taken to be equal to $2V_o$ (upper panel), $3V_o$ (in the middle) and $5V_o$ (bottom panel). Theoretical charge distributions are normalized to get total fragment yield equal to the measured one in the $Z$ range between 3 and 11. A remarkable density dependence of the calculated charge distributions is visible.

The least-square method has been used for quantitative comparison of the data and the calculations. Figure 2 shows the normalized $\chi^2$ as a function of $V/V_o$. From the minimum position and from the shape of the curve in its vicinity it is concluded that the best fit is obtained with the volume parameter $V_t = (2.9 \pm 0.2) V_o$. The error bar ($2\sigma$) is statistical one in origin. It is explained later why the subscript "t" is used.

### 3. SIZE OF EMITTING SOURCE

Generally, fragment kinetic energy is determined by four terms: thermal motion, Coulomb repulsion, rotation, and collective expansion, $E = E_{th} + E_C + E_{rot} + E_{flow}$. The Coulomb term is about three times larger than the thermal one [9]. The contributions of the rotational and flow energies are negligible for $p$+Au collisions [16]. Therefore the energy spectrum shape is essentially sensitive to the size of the emitting source. The kinetic energy spectra are obtained by calculation of multibody Coulomb trajectories, which starts with placing all charged particles of a given decay channel inside the freeze-out volume $V_f$. Each particle is assigned a thermal momentum corresponding to the channel temperature. The Coulomb trajectory calculations are performed for 3000 fm/$c$. After this amount of time the fragment kinetic energy is close to its asymptotic value. These calculations are the final step of the INC+Exp+SMM combined model.

Figure 3 gives a comparison of the measured (at $\theta = 87°$) carbon spectrum with the calculated ones (for emission polar angles $\theta = 87° \pm 7°$). The energy ranges of the spectra are restricted to 80 MeV to exclude the contribution of preequilibrium emission, which is possible at higher energies. The calculations are performed with a fixed volume at the partition moment, $V_t = 3V_o$, in accordance with the findings of the previous section. The freeze-out volume, $V_f$, is taken as a free parameter. Figure 3 shows the calculated spectra for $V_f / V_o$ equal to 3, 11, 19.

The least-square method is used to find the value of $V_f$ corresponding to the best description of the data. Figure 4 presents $\chi^2$ as a function of $V_f / V_o$. From the position of its minimum one gets $V_f = (11 \pm 3)V_o$. The systematics makes the main contribution to the error of this estimation of the freeze-out volume. It is caused by a 5% uncertainty in the energy scale calibration.

## 4. DISCUSSION

Existence of two different size parameters for multifragmentation has a transparent meaning. The first volume, $V_t$, corresponds to the partition point (or the moment of fragment formation), when the properly extended hot target spectator transforms into a configuration consisting of specified prefragments. These prefragments are not yet fully developed; there are still links (nuclear interaction) between them. The final channel of disintegration is completed during the dynamical evolution of the system up to the moment when receding and interacting prefragments become completely separated. This is just as in ordinary fission. The saddle point (which has a rather compact shape) resembles the final channel of fission by way of having a fairly well-defined mass asymmetry. Nuclear interaction between fission prefragments cease after descent of the system from the top of the fission barrier to the scission point. In papers by Lopez and Randrup [17,18] the similarity of both process was used to develop a theory of multifragmentation based on suitable generalization of the transition-state approximation first considered by Bohr and Wheeler for ordinary fission. The theory is able to calculate the potential energy as a function of the r.m.s. extension of the system yielding the space and energy characteristics of the transition configuration and barrier height for multifragmentation. The transition state is located at the top of the barrier or close to it. The phase space properties of the transition state are decisive for its further fate, for specifying the final channel.

Being conceptually similar to the approach of Ref. [17, 18], the statistical model of multifragmentation (SMM) uses the size parameters, which can be determined by fitting to data. The size parameter obtained from the IMF charge distribution can hardly be called a freeze-out volume. In the spirit of the papers by Lopez and Randrup we suggest the term "transition state volume", $V_t = (2.9 \pm 0.2) V_o$, therefore the subscript "t" is used.

The larger value of the size parameter obtained by the analysis of the kinetic energy spectra is a consequence of the main contribution of Coulomb repulsion to the IMF energy, which starts to work when the system has passed the " multi-scission point ". Thus, $V_f = (11 \pm 3) V_o$ is the freeze-out volume for multifragmentation in $p + Au$ collisions at 8.1GeV. It means that the nuclear interaction between fragments is still significant when the system volume is equal to $V_t$, and only when the system has expanded up to $V_f$, are the fragments freezing out.

In the statistical model used, the yield of a given final channel is proportional to the corresponding statistical weight. So, the nuclear interaction is neglected when the system volume is $V_t$, and this approach can be viewed as a rather simplified transition-state approximation. Nevertheless, the SMM describes well the IMF charge (mass) distributions for thermally driven multifragmentation. The point is clarified, if one looks through the recent paper [19], in which some particular approach (Dynamical Statistical Fragmentation Model) is developed to include the nuclear interaction in the calculation of the fragment yield. Comparison of the results of Ref. [19] with those obtained by the SMM reveals that both models give very similar shapes for the charge distributions of the lighter fragments considered in the present study.

Note that in the traditional application of the SMM only one size parameter is used, which is called "freeze-out volume". In the present paper we have demonstrated the shortcoming of such a simplification of the model. If only one parameter is used, one gets a significantly underestimated value of the freeze-out volume (half that given above).

## 5. SUMMARY

Analysis of the experimental data for target multifragmentation in $p(8.1\text{GeV}) + \text{Au}$ collisions results in the conclusion that within the framework of the statistical multifragmentation model there are two characteristic volume (or density) parameters. One, $V_t = (2.9 \pm 0.2)V_o$, corresponds to the configuration of the system at the moment of partition. It is similar to the saddle point in ordinary fission (transition state). Other, $V_f = (11 \pm 3)V_o$, is the freeze-out volume corresponding to the multi-scission point in terms of ordinary fission. The value of $V_t$ may be sensitive to the way of its estimation. The value of $V_f$ is extracted by means of multibody Coulomb trajectory calculations and should be only slightly model dependent. In further studies of the size or density parameters it is important to specify which stage of the system evolution is relevant to the observable chosen for the analysis.

**ACKNOWLEDGMENTS.** The authors are grateful to A. Hrynkiewicz, A.I. Malakhov, A.G. Olchevsky for support, to I.N. Mishustin, W. Reisdorf and other participants in the International Workshop Consensus Initiative (Catania, 2004) for illuminating discussions. The research was supported in part by the Russian Foundation for Basic Research, Grant № 03-02-17263, the Grant of the Polish Plenipotentiary to JINR, Bundesministerium für Forschung und Technologie, Contract № 06DA453, and the US National Science Foundation.**REFERENCES**

[1] J.P. Bondorf, A.S. Botvina, A.S. Iljinov, I.N. Mishustin and K. Sneppen, Phys. Rep. **257,** 133 (1995).

[2] D.H.E. Gross, Rep. Progr. Phys. **53,** 605 (1990).

[3] G. Sauer, H. Chandra and U. Mosel, Nucl. Phys. **A264,** 221 (1976).

[4] H. Jaqaman, A.Z. Mekjian and L. Zamick, Phys. Rev. C **27,** 2782 (1983).

[5] P.J. Siemens, Nature **305,** 410 (1983); Nucl. Phys. **A428,** 189c (1984).

[6] K.A. Bugaev, M.I. Gorenstein, I.N. Mishustin and W. Greiner, Phys. Rev. C **62,** 044320 (2000).

[7] V.E. Viola, Nucl. Phys. **A734,** 487 (2004).

[8] B. Bonderie, R. Bougault, P. Desesquelles et al., Nucl. Phys. **A734,** 495 (2004).

[9] V.K. Rodionov, S.P. Avdeyev, V.A. Karnaukhov et al., Nucl. Phys. **A700,** 457 (2002).

[10] V.A. Karnaukhov, H. Oeschler, S.P. Avdeyev et al., Phys.Rev. C **67,** 011601(R) (2003); Nucl. Phys. **A734,** 520 (2004).

[11] Bao-An Li, D.H.E. Gross, V. Lips and H. Oeschler, Phys. Lett. B **335,** 1 (1994).

[12] J.B. Natowitz, K. Hagel, Y. Ma et al., Phys. Rev. C **66,** 031601(R) (2002).

[13] M.D. Agostino, R. Bougault, F. Gulminelli et al., Nucl. Phys. **A699,** 795 (2002).

[14] V.D. Toneev, N.S. Amelin, K.K. Gudima and S.Yu. Sivoklokov, Nucl. Phys. **A519,** 463c (1990).

[15] W.A. Friedman, Phys.Rev. Lett. **60**, 2125 (1988).

[16] S.P Avdeyev, V.A. Karnaukhov, L.A. Petrov et al., Nucl.Phys. **A709,** 392 (2002).

[17] J.A. Lopez and J. Randrup, Nucl. Phys. **A503,** 183 (1989).

[18] J.A. Lopez and J. Randrup, Nucl. Phys. **A512,** 345 (1990).

[19] S.K. Samaddar and J.N. De, A. Bonasera, nucl-th/0402068v1 (20 Feb 2004).8

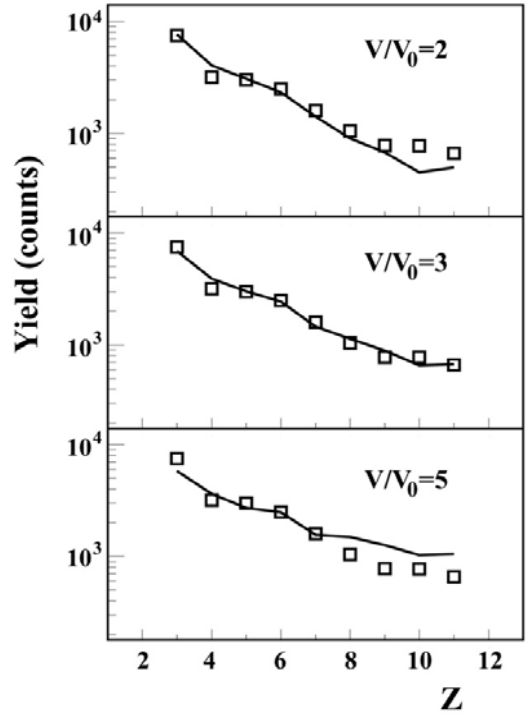

FIG. 1. Charge distribution of intermediate mass fragments measured for $p(8.1\text{GeV}) + \text{Au}$ collisions (dots) and calculated with the INC+Exp.+SMM prescription using different values of the system volume at the partition moment.

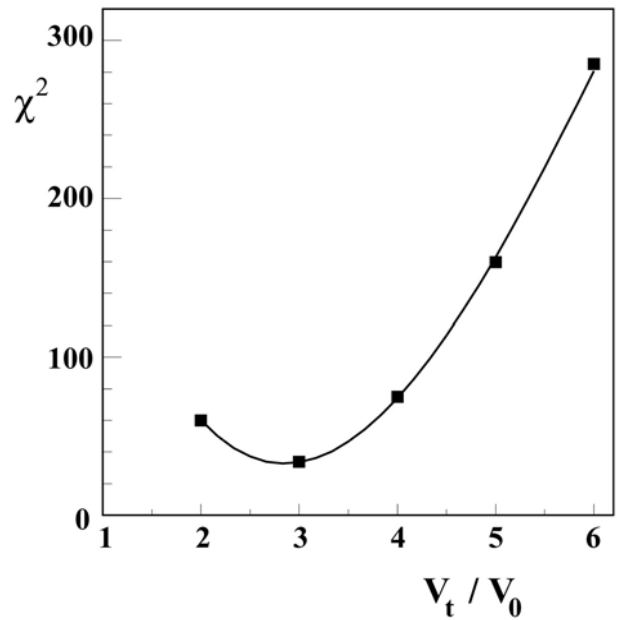

FIG. 2. Value of $\chi^2$ as a function of $V_t/V_o$ for comparison of the measured and calculated IMF charge distributions. The best fit of the model prediction to the data corresponds to $V_t = (2.9 \pm 0.2)V_o$.

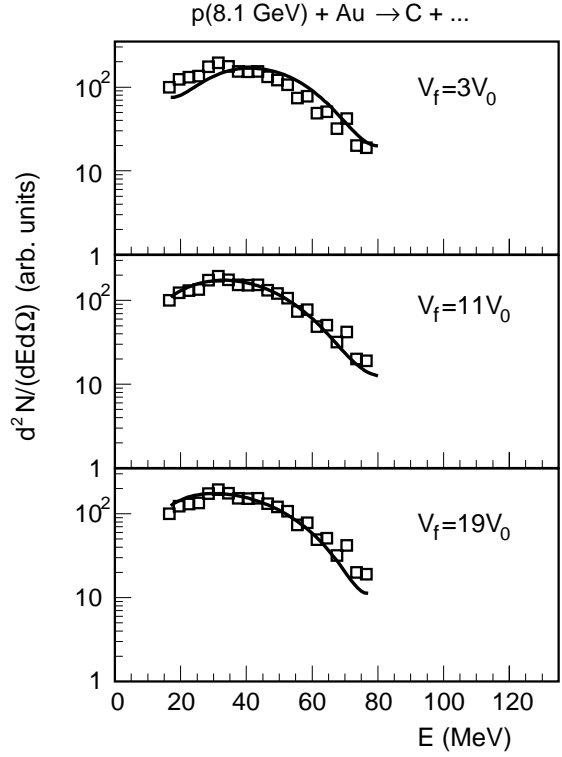

FIG. 3. Kinetic energy spectrum of carbon emitted (at $\theta = 87°$) by the target spectator in $p(8.1\text{GeV})+\text{Au}$ collisions. Symbols are the data, lines are calculated with the INC+Exp.+SMM prescription assuming the system volume at the partition moment $V_t = 3V_o$. The freeze-out volume, $V_f$, is taken to be equal to 3, 11, 19 $V_o$ (upper, middle, bottom panels).

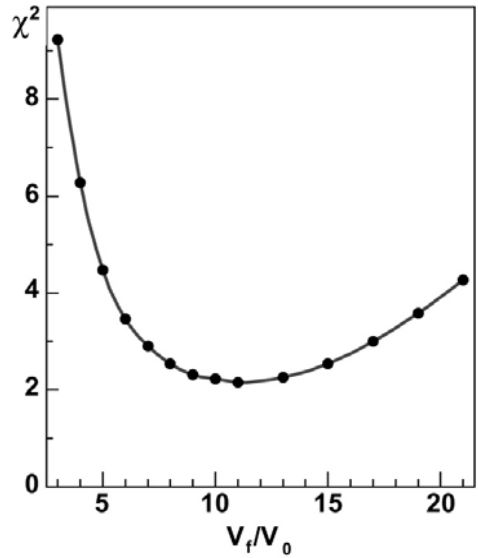

FIG. 4. Value of $\chi^2$ as a function of the freeze-out volume $V_f/V_o$ for comparison of the measured and calculated shapes of the carbon kinetic energy spectra, assuming a fixed size of the system at the partition moment, $V_t = 3V_o$. The best fit corresponds to $V_f = (11 \pm 3)\, V_o$.